\documentclass[11pt]{article}
\usepackage{epsfig}
\begin{document}

\def\wgta#1#2#3#4{\hbox{\rlap{\lower.35cm\hbox{$#1$}}
\hskip.2cm\rlap{\raise.25cm\hbox{$#2$}}
\rlap{\vrule width1.3cm height.4pt}
\hskip.55cm\rlap{\lower.6cm\hbox{\vrule width.4pt height1.2cm}}
\hskip.15cm
\rlap{\raise.25cm\hbox{$#3$}}\hskip.25cm\lower.35cm\hbox{$#4$}\hskip.6cm}}

\def\wgtb#1#2#3#4{\hbox{\rlap{\raise.25cm\hbox{$#2$}}
\hskip.2cm\rlap{\lower.35cm\hbox{$#1$}}
\rlap{\vrule width1.3cm height.4pt}
\hskip.55cm\rlap{\lower.6cm\hbox{\vrule width.4pt height1.2cm}}
\hskip.15cm
\rlap{\lower.35cm\hbox{$#4$}}\hskip.25cm\raise.25cm\hbox{$#3$}\hskip.6cm}}

\def\begeqar{\begin{eqnarray}}
\def\endeqar{\end{eqnarray}}

%
%
%
%

\title{Integrable quantum field theories with supergroup symmetries: 
the 
$OSP(1/2)$ case. }

\author{Hubert Saleur$^{a,b}$ and Birgit Wehefritz-Kaufmann$^{a,c}$\\
\smallskip\\
$^{a}$ Department of Physics and Astronomy\\
University of Southern California\\
Los Angeles, CA 90089\\
USA\\
\smallskip\\
$^{b}$ Service de Physique Th\'eorique\\
CEA Saclay\\
Gif Sur Yvette, 91191\\
France\\
\smallskip\\
$^{c}$ Physikalisches Institut der Universit\"at Bonn \\
Nu{\ss}allee 12\\
D-53115 Bonn\\
Germany}

\maketitle

\begin{abstract}
As a step to understand general patterns of 
integrability in $1+1$ quantum field theories 
with supergroup symmetry, we study in details the 
case of  $OSP(1/2)$. Our results include the solutions of natural
generalizations of models with ordinary group symmetry: the
$UOSP(1/2)_{k}$ WZW model with a current current perturbation,  
 the $UOSP(1/2)$ principal chiral model, and 
 the $UOSP(1/2)\otimes UOSP(1/2)/UOSP(1/2)$ coset models perturbed by
the adjoint. Graded parafermions are also discussed. A pattern peculiar 
to supergroups is the emergence of
another class of models, whose simplest representative is the
$OSP(1/2)/OSP(0/2)$ sigma model, where the  (non unitary) 
orthosymplectic symmetry is realized non linearly
(and can be spontaneously broken). For most models, we provide an integrable
lattice realization. We show in particular that integrable $osp(1/2)$
spin chains with integer spin flow to $UOSP(1/2)$ WZW models in the
continuum limit, hence providing what is to our knowledge the first
physical realization of a super WZW model.

\end{abstract}

\section{Introduction}

Two dimensional quantum field 
theories with supergroup symmetries have played an increasingly
important role in our attempts to understand phase transitions in 2D 
disordered systems - some recent works in this direction are
\cite{Denis, Chamon, Guruswamy, Fendley,Zirnbauerunpub,
Zirnbaueretal,Caux, BernardLeclairdis}. 

These theories  however prove quite 
difficult to tackle. Attempts at non perturbative approaches using 
conformal invariance \cite{Guruswamy,BernardLeclairdis}
 or exact S matrices \cite{SWKI,FendleyI,FendleyRead} have been popular recently, but so far, very few complete results are 
available. This paper is the second of 
a series (started with \cite{SWKI}) 
on models with orthosymplectic symmetry. Our goal is
to relate and 
identify the different pieces of the theoretical puzzle available -  
sigma models, Wess Zumino Witten (WZW) models 
and Gross Neveu (GN) models, integrable lattice models, and exactly 
factorized S matrices - and to find out which 
physical systems they describe, and which peculiarities arise from
the existence of supergroup symmetries. In our first
paper \cite{SWKI}, we  studied among other things the  $OSP(1/2)$ Gross Neveu model and the 
$OSP(1/2)/OSP(0/2)$ supersphere sigma model. A physical realization for 
the latter was identified in \cite{jacobsen}
in terms of a lattice loop model with self intersections, based on an
earlier work of \cite{nienhuis}. Other such realizations for
different models or supergroups have yet to be made. 
In the case of ordinary algebras, integrable lattice models do
provide such realizations, and 
are closely related 
with WZW and GN models 
based on the corresponding groups \cite{Affleck}.   
This relation is also important, for technical reasons,  in the solution of
the Principal Chiral Models (PCM) \cite{PW}. 

The main result of this paper is an analysis of integrable lattice
models based on the $osp(1/2)$ superalgebra, and the associated field
theories. While the general pattern is not unlike the case of
ordinary groups, important differences are also encountered. 

In section 2, we show that the continuum limit of the model based on 
the fundamental representation 
is not  the GN (or WZW model) but the supersphere sigma model, 
generalizing the observation of \cite{jacobsen}. 

In section 3 and 4 we show that 
that, for integer
spin, the continuum limit is the $UOSP(1/2)$ WZW model at integer level - 
in particular, the spin $1$ quantum spin chain flows to the 
$UOSP$ level one model. This
 provides, to our
knowledge, the 
first physical realization of a super WZW model. 
We also find  that for odd spin $s$, the continuum limit, like for $s=1$, is not
a WZW model. Attempts are made in section 6 to identify the
corresponding field theories, based on the expectation 
that in these  cases, the orthosymplectic symmetry is 
realized non linearly.

The $UOSP(1/2)$ PCM model is discussed in section 5, and the $UOSP(1/2)/U(1)$
models and associated parafermions in section 7.

\section{Integrable lattice models with $osp(1/2)$ symmetry }

Our conventions for the $osp(1/2)$ algebra  \cite{RittenbergScheunert} are summarized in the appendix . 
We start with the integrable model based on the fundamental representation $\rho_{1/2}$. The highest weight 
vector is denoted by $|1/2,1/2>$, and we shall treat it as  fermionic,
so the super dimension  of  this representation is equal to $-1$
\footnote{Changing the grading - that is treating the highest weight as bosonic - does not make the model into a `$O(1)$' model, and does not change any of the physical results. The grading we chose is simply more convenient, as it is well adapted to the structure of the symmetry algebra.}. The product of two
spin $1/2$ representations decomposes into a spin 0, a spin 1/2 and a
spin 1 representation. Their highest weights are respectively bosonic,
fermionic, and bosonic. The graded permutation operator reads 
\begin{equation}
P=-P_1+P_{1/2}+P_0
\end{equation}
and the Casimir 
\begin{equation}
2C=3P_1+P_{1/2}\equiv 3-2P_{1/2}-3P_0
\end{equation}

The hamiltonian of the integrable model is defined on the space $\rho_{1/2}^{\otimes N}$
as \cite{Kulish,Martins,tsuboi}
\begin{equation}
H=- c\sum_i {4\over 3}(P_0)_{i,i+1}+2(P_{1/2})_{i,i+1}
\end{equation}
(where $(P_j)_{i,i+1}$ denotes the projector onto spin $j$ in the
tensor product of the representations at site $i,i+1$, $c$ is a
normalization constant related with the sound velocity) 
is integrable, and corresponds to the anisotropic limit of the
integrable $osp(1/2)$ vertex model one can deduce from the scattering matrix of
\cite{SWKI}. The Bethe ansatz equations for this model read schematically
\begin{equation}
\left({\lambda-i/2\over\lambda+i/2}\right)^N=\epsilon\prod
{\lambda-\lambda'-i\over \lambda-\lambda'+i}\prod
{\lambda-\lambda'+i/2\over \lambda-\lambda'-i/2}\label{bethe}
\end{equation}
(where the $\lambda$'s are the roots) and the energy 
\begin{equation}
E=-c\sum{1\over \lambda^2+1/4}
\end{equation}
The sign $\epsilon$ depends on the boundary conditions for the
hamiltonian, and has not, in our opinion, always been correctly interpreted
in the literature 
\cite{Martins}. The point is that a hamiltonian with $osp(1/2)$
symmetry will be obtained by having the last term in the sum involve
the projectors $(P_j)_{N,N+1}$, and identifying the states in the
$N+1^{th}$ space with the ones in the first space. In the case of superalgebras, this is not exactly
the same as having the projectors $(P_j)_{N,1}$: the difference
involves `passing generators' through the $N$ first states in the tensor
product, and this can of course generate signs. 
The hamiltonian with $osp(1/2)$ symmetry corresponds to the Bethe
equations with $\epsilon=1$ in (\ref{bethe}). This agrees with the original results in \cite{Kulish}.  Antiperiodic boundary conditions for the
fermions
would correspond to $\epsilon=-1$ instead.

According to Martins \cite{Martins}, when $\epsilon=1$, 
the ground state of the $S^z=0$ 
coincides with the one of the $S^z=1/2$ 
sector ,
leading to a degeneracy of 4 for the state $h=\bar{h}=0$. The central 
charge read in that sector
is $c=-2$. The total partition function (that is, the 
trace of $q^{(H+P)/2}\bar{q}^{(H-P)/2}$, $P$ the momentum, and 
for $\epsilon=1$ again) reads from
 \cite{Martins}
\begin{eqnarray}
Z&=&4\left|q^{1/12}\prod_{n=1}^\infty (1+q^n)^2\right|^2\nonumber\\
&=&{1\over \eta\bar{\eta}}\sum_{m\in 2Z,e\in Z+1/2} q^{(2e+m)^2/8}
\bar{q}^{(2e-m)^2/8}\label{parfun}
\end{eqnarray}
This is  in agreement with the interpretation of the low energy limit  of
this lattice model with a  symplectic fermion theory, as was proposed
in \cite{jacobsen}. In the latter paper, this identification was made
by using the fact that the hamiltonian is the anisotropic limit of a
vertex model which can be reinterpreted as a loop model, and thus as a
model of classical $OSP(1/2)$ spins in two dimensions, similar to the
one used in the analysis of the usual $O(n)$ model. It was then argued
that the integrable hamiltonian lies in the broken symmetry Goldstone
phase,
and that the low energy limit is 
the weak coupling limit of the supersphere sigma model,
whose target space is $S^{(0,2)}=OSP(1/2)/[OSP(0/2)\equiv SP(2)]$ (the equivalent of
$O(N)/O(N-1)$ for $N=-1$). Recall one can easily parametrize this target space 
using $x=1-\eta_{1}\eta_{2}$ such that $x^{2}+2\eta_{1}\eta_{2}=1$. The sigma model action (Boltzmann weight $e^{-S}$)
is 
\begin{equation}
S={1\over g}\int d^2 x\left[ (\partial_{\mu}x)^{2}+
 2\partial_{\mu}\eta_{1}
\partial_{\mu}\eta_{2}\right]\label{basicsigma}
\end{equation}
with the beta function $\beta\propto -3g^{2}$. At small coupling, the action reduces to the symplectic fermions 
theory, and the partition function (\ref{parfun}) coincides with
the determinant of  the Laplacian with periodic boundary conditions 
in the space direction and antiperiodic boundary conditions in 
the ``time'' direction (along which the trace is taken). For $g$ negative,
the model flows to weak coupling in the UV, and is massive in the IR, where symmetry is 
restored. The action reads then, in terms of the fermion variables,
and after trivial rescalings,
\begin{equation}
    S=-{1\over |g|}\int d^2
   x\left[\partial_{\mu}\eta_{1}\partial_{\mu}\eta_{2}-\eta_{1}\eta_{2}
    \partial_{\mu}\eta_{1}\partial_{\mu}\eta_{2}\right]
    \end{equation}
Notice that the relative normalization of the two terms can be
changed at will by changing the normalization of the fermions. The
relative sign can also be changed by switching the fermion labels
$1\rightarrow 2$. However, the sign of the four fermion term cannot
be changed, and determines whether the model is massive or massless
in the IR. For 
$g$ positive, the model flows (perturbatively) to weak coupling in the IR. This is the
case of the lattice model introduced in \cite{nienhuis,Martins}.

It is possible to generalize the integrable model by introducing
heterogeneities in a way well understood for  ordinary
algebras \cite{reshetikhin}. In doing so, the source term in the equations
(\ref{bethe})  is replaced 
by 
\begin{equation}
\left({\lambda-\Lambda-i/2\over \lambda-\Lambda+i/2}\right)^{N/2}
\left({\lambda+\Lambda-i/2\over \lambda+\Lambda+i/2}\right)^{N/2}
\end{equation}
where $\Lambda$ is a parameter measuring heterogeneities,
and the energy becomes 
\begin{equation}
E=-{c\over 2} \sum {1\over (\lambda-\Lambda)^2+1/4}+{1\over
  (\lambda+\Lambda)^2+1/4}
\end{equation}
We will not discuss complete calculations here, but simply  derive some essential
features of the associated thermodynamics Bethe ansatz (TBA). The ground state is made of real particles, and excitations are holes in the ground state. After introducing the Frourier transforms
\begin{equation}
\hat{f}(x)=\int d\lambda e^{i\lambda x}f(\lambda),~~f(\lambda)={1\over
  2\pi}
\int dx e^{-i\lambda x}\hat{f}(x)
\end{equation}
the
physical equations read
\begin{equation}
\hat{\rho}+\hat{\rho}^h= {\cos \Lambda x\over
  2\cosh(x/2)-1}
- e^{-|x|/4}{\sinh (x/2)\over \cosh (3x/4)} \hat{\rho}^h
\end{equation}
and the energy, up to a constant
\begin{equation}
E={c\over 2\pi} \int \hat{\rho}(x){\cos\Lambda x\over 2\cosh(x/2)-1}dx
\end{equation}
The interesting way to proceed then is to take the limit
$N\rightarrow\infty$, $a\rightarrow 0$ ($a$ the lattice spacing), such that $Na\rightarrow L$
finite. We then take the limit $\Lambda
\rightarrow\infty$ with $e^{-2\Lambda\pi/3}/a$ finite. In that limit,
excitations at finite rapidity acquire a relativistic dispersion
relation, with rapidity $\theta={2\pi\over 3}\lambda$. The scattering
of these excitations with themselves corresponds to the $S$ matrix
element:
\begin{equation}
S\equiv\Sigma_0=-\exp\left[i\int_{-\infty}^\infty
  {d\omega\over\omega}e^{-3i\omega/\pi}e^{-|\omega|/2}
  {\sinh\omega\over\cosh(3\omega/2)}\right]\label{sigma0elt}
\end{equation}
and the latter coincides with $\sigma_3^+-\sigma_2^+$, the scattering
matrix element of particle 1 with itself in the sigma model (\ref{basicsigma}), as discussed in 
\cite{SWKI} (this matrix element is called $\Sigma_0$ there)\footnote{
Misprints have unfortunately cropped up in the equation whose
denominator should read $\sinh\omega\cosh(\omega(3\xi-\pi)/2\pi)$
instead.}.

 In fact, one can check that
the thermodynamics of the spin chain, in this limit, coincides with
the thermodynamics of the field theory for the supersphere sigma model
discussed in \cite{SWKI}: the introduction of heterogeneities
provides thus a regularization of this field theory. 

As always - and this can be related \cite{destri} to  the 
Nielsen-Ninomiya theorem \cite{nielsen} -  the massive degrees of freedom
near vanishing bare rapidity in the model with heterogenities are completed by massless degrees of freedom at large bare rapidities 
(edges of the Brillouin zone). These are the same massless modes that would be 
present in the homogeneous chain obtained by letting $\Lambda=0$. The dynamics of these
massless modes decouples entirely from the dynamics of the massive ones, 
and one can identify the associated CFT with the weak coupling limit of the 
supersphere sigma model, that is, the symplectic fermion theory.

It is tempting to carry out the same procedure for the case of higher
spin. Unfortunately, not much is known about the higher spin
integrable $osp$ spin chains in explicit form. It is fair to expect, based on analogies
with other cases - in particular the $so(n)$ case - that such chains do
exist, and are described by changing the source terms and energy terms
as  
\begin{equation}
{\lambda -i/2\over\lambda+i/2}\rightarrow {\lambda-si\over
  \lambda+si},~~~
{1\over\lambda^2+1/4}\rightarrow {2s\over\lambda^2+s^2} 
\end{equation}
where $s$ is the higher spin. The thermodynamics of the massive  field
theory limit is described by the equations 
\begin{equation}
    {\epsilon_j(\theta)\over T}=\phi(\theta-\theta')*
    \ln\left(1+e^{\epsilon_j(\theta')/T}\right)-
    \sum_{l=1}^\infty  \left(\delta_{j,l+1}+\delta_{j,l-1}\right)
\phi(\theta-\theta')*
    \ln\left(1+e^{-\epsilon_l(\theta')/T}\right)
\end{equation}
where $\phi(\theta)={3\over 2\cosh 3\theta/2}$ and $f*g(\theta)={1\over 2\pi}
\int f(\theta-\theta')g(\theta')d\theta'$. The boundary condition $\epsilon_{2s}\rightarrow
m\cosh\theta$ must be imposed. The free energy reads then
\begin{equation}
    F=-T\int {d\theta\over 2\pi} m\cosh\theta\ln\left(1+e^{-\epsilon_{2s}/T}\right)
    \end{equation}

The thermodynamics of the lattice model is described by similar equations, but different source terms. It allows one
 in particular to determine the entropy per site of the chain in the large $T$ limit. One finds 
 that this entropy corresponds, for $s$ half integer, to a mix of representations
 $\rho_{1/2},\rho_{3/2},\ldots,\rho_{s}$, and for $s$ integer, a mix of 
 representations $\rho_{0},\rho_{1},\ldots,\rho_{s}$. The integrable models
 must therefore involve these mix of representations on every site, and 
 presumably  must be considered as having 
 $osp$ super-Yangian symmetry, 
 in analogy with the $so(n)$ case \cite{McKay}. In particular, the
extension of  the adjoint by a scalar representation to form
an irreducible  representation of the Yangian is typical.  
Calculations with a
 twist angle giving antiperiodic boundary conditions to the kinks
\footnote{This is analogous to the study of excited states carried out
 in \cite{tsuboi}. } shows that the representations with half-integer spin have
 superdimension $-1$, while those with integer spin have
 superdimension $+1$.  
Some of these results have been obtained independently and using a
different approach in \cite{tsuboi}.

It is easy to check that the central charge of these models is 
 \begin{equation}
     c_{eff}={8s\over 2s+3}\label{firstceff}
     \end{equation}

As in the usual $su(2)$ case, one can deform the models by considering 
$R$ matrices with $U_qosp(1/2)$ symmetry, and one can truncate them in 
the case $q$ a root of unity. The resulting TBA's have the form shown in
Figure 1 (with a total number of nodes equal to $N$), and central charge
\begin{equation}
c_{eff}={8s\over 2s+3}-{24s\over (N+4)(N+4-2s)}\label{secondceff}
\end{equation}
\begin{figure}
\begin{center}
\epsfig{file=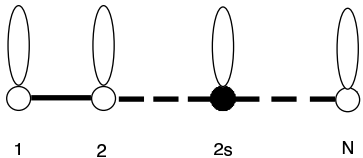,scale=0.5}  
\caption{\label{fig0} Incidence diagram of the general TBA obtained
after quantum group deformation and truncation of the spin $s$ chain.}
\end{center}
\end{figure}

Most of the following 
is devoted to understanding the field theories associated with
(\ref{firstceff}) and (\ref{secondceff}).

\section{Coset models}

The basic field theory we have introduced so far is the $OSP(1/2)/SP(2)$ non linear sigma model (\ref{basicsigma}). Another type of sigma model plays a major role in the analysis: the 
$UOSP(1/2)_k$ Wess Zumino Witten model. Details about $OSP$ and $UOSP$ are furnished in the appendix:  
the
bosonic part of $UOSP(1/2)$  is $SU(2)$, and the group is compact. The
level $k$ is quantized (for the
normalization of $k$, we use the level of the sub $SU(2)$, like for instance
in the works \cite{ospwzw}. The same model  would be  called 
the $OSP(1/2)_{-2k}$ model
following the conventions used in the literature on disordered systems
(see eg \cite{andreas}, as well
as in our previous paper).  The model is not expected to be a unitary
conformal field theory: this is clear at the level of the action,
where for instance the purely fermionic part is closely related to the
$\eta\xi$ system, a non unitary theory. This is also expected on
general grounds, since, for instance, there is no way to define a
metric without negative norm (square)  states in some representations.

It turns out however that the $UOSP(1/2)_k$ WZW  theories are relatively simple, at
least at first sight. The best way to understand them is to 
use a remarkable  embedding discovered by Fan
and Yu \cite{embedding}.

\subsection{The $UOSP(1/2)/SU(2)$ coset models}

These authors made the crucial observation that 
\begin{equation}
UOSP(1,2)_k\approx SU(2)_k \times {UOSP(1,2)_k\over SU(2)_k}
\end{equation}
where the branching functions of the latter part define a Virasoro
minimal model, with 
\begin{eqnarray}
c_{uosp}&=&{2k\over 2k+3}\nonumber\\
c_{su2}&=&{3k\over k+2}\nonumber\\
c_{virasoro}&=&1-6{(k+1)^2\over (k+2)(2k+3)}
\end{eqnarray}

Only for $k$ an integer does the action of the Wess Zumino model make 
sense,
and we will restrict ourselves to this case in the following.
 The
Virasoro models which appear there have $p=2k+3,q=k+2$; they are non
unitary, 
and their effective central charge is $c_{eff}=1-{12\over
  (2k+3)(2k+4)}$. These models can thus be considered as $UOSP/SU$
coset models! 

The  perturbation of these models  by the
operator $\phi_{21}$ (here, the labels refer to the description as a Virasoro minimal model) with dimension $h=1-{3\over 4(k+2)}$  is well
known to be integrable (the
$1$ comes from the $OSP$, the $3/4(k+2)$ from the $SU(2)$). The TBA has the form  shown in  
Figure 2 \cite{ravanini}. As observed in \cite{SWKI}, it can be obtained 
after a q-deformation and a truncation of the basic supersphere sigma model TBA.
The 
corresponding S matrices can thus easily  be deduced, and follow 
RSOS restrictions of the q-deformed $a_2^{(2)}$ S matrices, or,
equivalently,  q-deformed $osp(1/2)^{(1)}$ S matrices. The
simplest and most interesting case corresponds to the model of
Virasoro minimal series $p=5,q=3$. Its central charge is $c=-3/5$
while $c_{eff}=3/5$. The TBA for a perturbation by the operator
$\phi_{21}$ of weight $h=3/4$ is described by the diagram in the
figure
in the particular case where the number of nodes is two. The S matrix 
has
been worked out in details in \cite{Mussardo}. 

\begin{figure}
\begin{center}
\epsfig{file=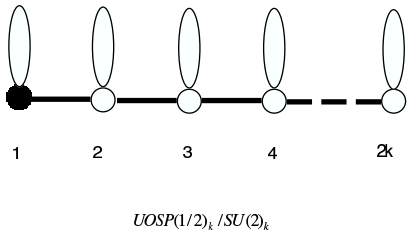,scale=0.5}  
\caption{\label{fig1} Incidence diagram of the TBA describing 
$UOSP(1/2)_k/SU(2)_k$ coset
  models perturbed by  the operator with $h=1-{3\over 4(k+2)}$. The
  total number of nodes is $2k$.}
\end{center}
\end{figure}

An amusing  consequence of this observation  is that  the supersphere 
sigma model
appears as the limit $k\rightarrow\infty$ of a series of coset
models. This is quite similar to the way  the ordinary sphere sigma 
model appears as the limit of
a series of parafermion theories  \cite{FateevZamo}, this time of  type
$SU(2)_{k}/U(1)$. 

An important difference between the two cases 
 is that, since the three point function of $\phi_{21}$ vanishes, the
perturbation of the coset models is independant of the sign of the coupling, and thus always massive. The situation was different
 in the case of parafermionic theories $SU(2)/U(1)$, where one
 sign was massive (and corresponded, in the limit
 $k\rightarrow\infty$, to the case $\theta=0$), but the other was
 massless \cite{FateevZamo} (and corresponded in the limit $k\rightarrow\infty$, to the
 case $\theta=\pi$). For the supersphere, there is no theta term, so 
 it is natural that we get only one flow. 
\footnote{Recall that $\Pi_2(S^{m-1/2n})=\Pi_2(S^{m-1})=0$ for $m\neq
 3$, $=Z$ for $m=3$.}

An interesting consequence of the embedding  is that we can 
deduce the effective
central charge of the $UOSP(1/2)$ WZW model at level $k$. Using that 
for the Virasoro model, $c_{eff}=1-{12\over (2k+2)(2k+3)}$, one finds 
\begin{equation}
c_{eff}={8k\over 2k+3},~~~UOSP(1/2)_k\label{ceff}
\end{equation}
This result will be compatible with all the subsequent analysis, but it 
is in slight disagreement with \cite{ospwzw,embedding}. In the latter papers, 
conjectures are made that the spectrum closes on primary fields of spin $j=0,{1\over 2}
,\ldots,{k\over 2}$ with dimension $h={j(2j+1)\over 2k+3}$. If
this turned out to be true, the models we identify would not
exactly be the WZW models, but maybe some ``extensions'' of 
these - at the present time, this issue is not settled, but
it seems simpler to assume the value  (\ref{ceff}) is indeed the 
effective central charge of the WZW model.

\subsection{The $UOSP(1/2)\times SU(2)/SU(2)$ coset models}

We consider now TBA's with a total number of nodes $N=2k+2l$. If the massive node is the $2k^{th}$ one, the UV central charge is 
\begin{eqnarray}
c_{eff}&=&{8k\over 2k+3}-{24k\over (2k+2l+4)(2l+4)}\nonumber\\
&=&{8k\over 2k+3}+{3l\over l+2}-{3(l+k)\over l+k+2}
\end{eqnarray}
suggesting that the model can be understood as a coset model $UOSP(1/2)_k\otimes SU(2)_l/SU(2)_{k+l}$. Assuming the TBA corresponds to a theory perturbed by an operator whose odd point functions vanish, we find the dimension of the 
 perturbing operator to  be  $h=1-{3\over 4(k+l+2)}$. This is compatible with
taking  the spin $1/2$ field in the denominator of the coset.

\begin{figure}
\begin{center}
\epsfig{file=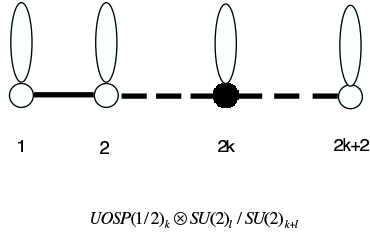,scale=0.5}  
\caption{\label{fig2} Incidence diagram of the TBA describing 
$UOSP(1/2)_k\otimes SU(2)_l/SU(2)_{k+l}$ coset
  models perturbed by  the operator with $h=1-{3\over 4(k+l+2)}$. The
  total number of nodes is $2k+2l$.}
\end{center}
\end{figure}

If the massive node is the $2k+1^{th}$ one meanwhile, the central charge is
\begin{eqnarray}
c_{eff}&=&{8k+4\over 2k+4}-{12(2k+1)\over (2k+2l+4)(2l+3)}\nonumber\\
&=& {3k\over k+2}+{8l\over 2l+3}-{3(k+l)\over k+l+2}
\end{eqnarray}
suggesting similarly that the model can be understood as a coset $SU(2)_k\otimes UOSP(1/2)_l/SU(2)_{k+l}$ perturbed by the operator of dimension $h=1-{3\over 4(k+l+2)}$. Of course the two cases are actually equivalent by taking mirror images, but it is convenient to keep them separate to study the large $l$ limit later. 

\begin{figure}
\begin{center}
\epsfig{file=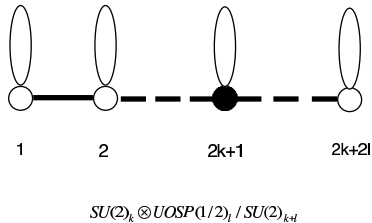,scale=0.5}  
\caption{\label{fig3} Incidence diagram of the TBA describing 
$SU(2)_k\otimes UOSP(1/2)_l/SU(2)_{k+l}$ coset
  models perturbed by  the operator with $h=1-{3\over 4(k+l+2)}$. The
  total number of nodes is $2k+2l$.}
\end{center}
\end{figure}

\subsection{The $UOSP(1/2)\otimes UOSP(1/2)/UOSP(1/2)$ models.}

We now consider instead TBA's with a total number of nodes $N=2k+2l-1$. If the massive node is the $2k^{th}$ one, the UV central charge is found to be
\begin{eqnarray}
c_{eff}&=&{8k\over 2k+3}-{24k\over(2l+3)(2k+2l+3)}\nonumber\\
&=& {8k\over 2k+3}+{8l\over 2l+3}-{8(k+l)\over 2k+2l+3}
\end{eqnarray}
suggesting that the models can be interpreted as coset 
$UOSP(1/2)_k\otimes UOSP(1/2)_l/UOSP(1/2)_{k+l}$. Assuming 
the TBA corresponds to a theory perturbed by an operator whose 
odd point functions do not vanish, we find the dimension of the perturbing operator to be 
$h=1-{3\over 2k+2l+3)}$. This is compatible with taking 
the spin $1/2$ field in the denominator of the coset. 

\begin{figure}
\begin{center}
\epsfig{file=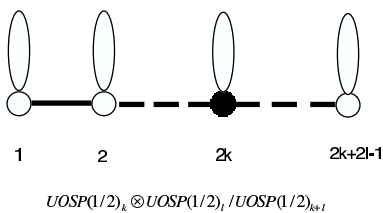,scale=0.5}  
\caption{\label{fig4} Incidence diagram of the TBA describing 
$UOSP(1/2)_k\otimes UOSP(1/2)_l/UOSP(1/2)_{k+l}$ coset
  models perturbed by  the operator with $h=1-{3\over 2k+2l+3}$. The
  total number of nodes is $2k+2l-1$.}
\end{center}
\end{figure}

Note that, since we have assumed the three point function of the
perturbing operator does not vanish, switching the sign of the
perturbation should lead to a different result. It is 
natural to expect that one has then a massless flow, whose TBA and S
matrices are readily built by analogy with the $SU(2)$ case
\cite{Zamomassless}: we leave
this to the reader as an exercise. 

Finally, we notice that the  $UOSP(1/2)$ coset model with $k=l=1$   was first identified in the paper \cite{Dorey}.

\subsection{The other models}

The last possible case we can obtain out of this construction
corresponds to a TBA's with an odd number of nodes 
(say, $2k+1$), and the mass on an odd node, too.

The  effective central charge is
$c_{eff}=1-{12\over (2k+5)(2k+4)}$. The models can be considered as
Virasoro models with  $p=2k+5,q=k+2$, and the TBA corresponds to
perturbation 
by the $\phi_{15}$ field now, of dimension $h_{15}=1-{3\over
  2k+5}$.  We have not found any convincing
way to interpret this in terms of $OSP(1/2)$ cosets; maybe it is not 
possible.  
Notice that the $3/(2k+5)$ is a weight for
$OSP_{k+1}$, which, since it appears with a minus sign in $h$, should
be in the denominator of the sought after coset. 
Notice also that, by using the remark at the end of the previous
paragraph, 
we expect flows between the  models we have interpreted in terms of
$OSP(1/2)$ and $SU(2)$ cosets and these unidentified models. This
could be a useful hint.

\section{Sigma models}

\subsection{The $UOSP$ WZW models}

Taking $l\rightarrow\infty$ for the class of models where the massive node is an even one, we obtain theories with central charge $c_{eff}={8k\over 2k+3}$. 
This value coincides with the result obtained in the first section for $s=k$. We therefore suggest that 
the continuum limit of the
lattice models with {\sl integer spin} $s$ are the $UOSP(1/2)_{k=s}$ models. Introducing heterogeneities 
then gives rise to the current-current perturbation of these models.

\begin{figure}
\begin{center}
\epsfig{file=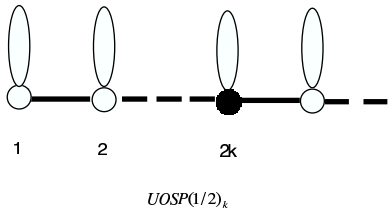,scale=0.5}  
\caption{\label{fig5} Incidence diagram of the TBA describing the $UOSP(1/2)_k$ WZW model with a current current perturbation.}
\end{center}
\end{figure}

The S matrix is the tensor product of 
the RSOS S matrix for the
Virasoro model $M_{2k+3,k+2}$ perturbed by $\phi_{21}$ (which we saw
can be reinterpreted as an $UOSP$ RSOS matrix) and 
the supersphere sigma model S matrix.

These results
apply to the NS sector of the model, where the fermionic currents 
have integer modes,
and are periodic. The Ramond sector can be obtained by spectral flow; 
one has in particular \cite{embedding}
\begin{equation}
L_0^{R}=L_{0}^{NS}-J_{3}^{0,NS}+{k\over 4}\label{spectralflow}
\end{equation}
While the true central charge seems inaccessible from the TBA,
one can follow the spectral flow by giving a fugacity to the 
solitons, 
as was discussed in our first paper, ie calculating 
$Z=\hbox{Tr} \left [e^{-\beta H}e^{i\alpha q/(t-1)}\right]$,
where $q$ is the topological charge of the solitons, normalized as $q=0,\pm 1$. 
Antiperiodic boundary conditions correspond to $\alpha=(t-1)\pi$, and 
 are found to give, using the system of equations (38,39) of our previous paper
\begin{equation}
c_{{eff}}={8k\over 2k+3}-6k
\end{equation}
in agreement with  (\ref{spectralflow}). 

Finally, it is easy to check from the TBA that the dimension of the perturbing operator has to be $(1,1)$. 
This gives strong support to our conjecture. 

We stress that, as far as we know,  none of the perturbed $UOSP(1/2)_k$ 
WZW  models can be interpreted as a
Gross-Neveu model. The $OSP$ GN models correspond to models with,
formally, level $k=-{1\over 2}$, and have a different physics, and
different scattering matrices, as discussed in \cite{SWKI}. We will 
get back to this issue in the conclusion.


\subsection{The ``$SU(2)_k\otimes UOSP(1/2)/SU(2)$'' models.}

If we take the limit $l\rightarrow\infty$ for models which have 
the mass on an odd node, the central charge as well as 
the interpretation of the coset models are consistent 
with a theory of the form $SU(2)_k\otimes UOSP(1/2)/SU(2)$,
of which the supersphere sigma model was just the simplest ($k=0$) version. 

\begin{figure}
\begin{center}
\epsfig{file=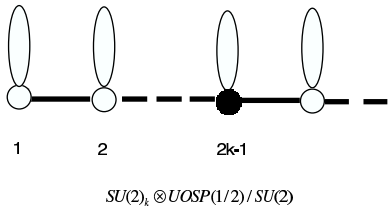,scale=0.5}  
\caption{\label{fig6} Incidence diagram of the TBA describing the $SU(2)_k
\otimes UOSP(1/2)/SU(2)$ sigma model.}
\end{center}
\end{figure}

It would be most interesting to find out the
action describing these models, but we have not done so
for now - we will comment about the problem below.

\section{The $UOSP(1/2)$ PCM model}

In the $SU(2)$ case for instance, the limit $k\rightarrow\infty$ of the WZW model 
with a current current perturbation coincides with the 
scattering theory for the PCM (principal chiral model) model \cite{PW}. It is natural to expect that the same thing will
hold for the $UOSP(1/2)$ case. The TBA looks as in
Figure 8, and the  
   scattering matrix has obviously the form 
$S_{PCM}\propto S\otimes S$, 
where $S$ is the S matrix for the supersphere sigma model, up to
CDD factors we will discuss below 

\begin{figure}
\begin{center}
\epsfig{file=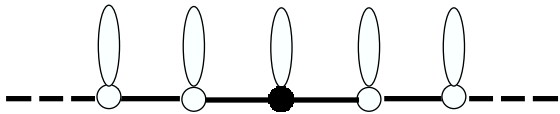,scale=0.5}  
\caption{\label{fig7} Incidence diagram of the TBA describing the 
$OSP(1/2)$ PCM model.}
\end{center}
\end{figure}

Let us study  this PCM model more explicitely. 
It is convenient to write an element of $UOSP(1/2)$ as 
\begin{equation}
g=\left(\begin{array}{ccc}
1+{1\over 4}\eta\eta^\diamond&-{1\over 2}\eta&{1\over 
2}\eta^\diamond\\
-{1\over 2}(a\eta^\diamond-b^\diamond\eta)&a(1-{1\over
  8}\eta\eta^\diamond)&-
b^\diamond(1-{1\over 8}\eta\eta^\diamond)\\
-{1\over 2}(b\eta^\diamond+a^\diamond\eta)&b(1-{1\over
  8}\eta\eta^\diamond)& 
a^\diamond(1-{1\over 8}\eta\eta^\diamond)
\end{array}\right)
\end{equation}
with the constraint $aa^\diamond+bb^\diamond=1$.
In a similar way, the conjugate of the matrix, $g^\ddag$, reads
\begin{equation}
g^\ddag=\left(\begin{array}{ccc}
1+{1\over 4}\eta\eta^\diamond&{1\over 
2}(b\eta^\diamond+a^\diamond\eta)
&-{1\over 2}(a\eta^\diamond-b^\diamond\eta\\
{1\over 2}\eta^\diamond&a^\diamond(1-{1\over
  8}\eta\eta^\diamond)&
b^\diamond(1-{1\over 8}\eta\eta^\diamond)\\
{1\over 2}\eta&-b(1-{1\over
  8}\eta\eta^\diamond)& 
a(1-{1\over 8}\eta\eta^\diamond)
\end{array}\right)
\end{equation}

The action of the PCM model reads, after a rescaling of the
fermions $\eta\rightarrow 2\eta$
\begin{eqnarray}
-Str \left(\partial_\mu g\partial_\mu g^\dagger\right)\propto 
\partial_\mu\eta\partial_\mu\eta^\diamond +
\left(\partial_\mu
  a\partial_\mu a^\diamond+\partial_\mu b\partial_\mu
  b^\diamond\right)\left(1-\eta\eta^\diamond\right)+
{1\over 2}
\eta\eta^\diamond\partial_\mu\eta\partial_\mu\eta^\diamond\label{eqI}
\end{eqnarray}

We note that the $UOSP(1/2)$  group
manifold can be identified with the supersphere
$S^{3,2}$\cite{Landi}, that is, the space 
 $OSP(4/2)/OSP(3/2)$. The PCM model, however, cannot be expected to 
 coincide with the sigma model on $S^{3,2}$:
 the symmetry groups are different, and so are the invariant actions.
For instance, in the PCM model, the group $UOSP(1/2)$ 
 acts by conjugation, leaving the identity invariant. In the vicinity
 of the identity, under the $SP(2)=SU(2)$, the fermionic 
coordinates transform as a doublet, and the bosonic coordinates transform as a
 triplet. 
 In the sigma model, the coordinates near the origin transform as the
 fundamental of $OSP(3/2)$. Under the $SO(3)=SP(2)=SU(2)$ of the $OSP(3/2)$, 
the bosonic coordinates transform as a triplet
{\sl but} the
 fermionic coordinates now transform as a singlet (they form a doublet
 under a different $SP(2)$, which leaves the sphere $S^3$ invariant). The groups acting differently, the invariant actions can be expected to be  different.  
 This is confirmed by explicit calculation.
 The supersphere $S^{3,2}$ can be parametrized in terms of
coordinates $x_i$, $i=0,\ldots,3$ and $\eta_1,\eta_2$. The 
constraint  $\sum_0^3
x_i^2+2\eta_1\eta_2=1$ gives rise to
\begin{equation}
x_i=y_i\left(1-\eta_1\eta_2\right),~~\sum_0^3 y_i^2=1
\end{equation}
The sigma model action 
\begin{equation}
S=2\partial_\mu\eta_1\partial_\mu\eta_2+\sum_{i=0}^3 (\partial_\mu x_i)^2\label{eqII}
\end{equation}
becomes then
\begin{equation}
S=2\partial_\mu\eta_1\partial_\mu\eta_2
+
\left(\sum_{i=0}^3
  \left(\partial_\mu y_i\right)^2\right)
  (1-2\eta_1\eta_2)-2\eta_1\eta_2\partial_\mu\eta_1\partial_\mu\eta_2
\end{equation}
The two equations (\ref{eqI},\ref{eqII}) are similar, but exhibit a
major difference in 
the sign of the four fermion term. 

The  physics of the two models  is 
considerably different. For the supersphere sigma model,
the $\beta$ function is exactly zero to all orders, and the theory
 is exactly conformal invariant for any value of the coupling 
constant (like in the $O(2)/O(1)$ case).  For the PCM, the $\beta$ 
function
follows from Wegner's calculations in the case $O(-1)$ \cite{Wegner} 
\begin{equation}
\beta=3\lambda^2-{9\over 2}\lambda^3+{81\over 8}\lambda^{4}
\ldots
\end{equation}
to be compared eg with the $SU(2)$ case
\begin{equation}
\beta=-2\lambda^2-2\lambda^3-3\lambda^4+\ldots
\end{equation}
The conventions here are that the Boltzmann weight is 
$\exp\left(- S\right)$, and 
$$
S=-{1\over 2\lambda}\int Tr~(Str~)\left[\partial_\mu g\partial_\mu 
g^\dagger\right]={1\over 2\lambda} \int Tr \left[g^{-1}\partial_\mu g\right]^2
$$
In the $SU(2)$ case, the massive theory corresponds to $\lambda<0$. By contrast, for the $OSP(1/2)$ case, the massive direction 
corresponds to $\lambda>0$. However, since one takes then a supertrace instead of a trace, the $SU(2)$ part of the PCM 
action has  the {\sl same} sign as in the $SU(2)$ pure case, with 
Boltzmann weight 
$\exp[-|cst|\int (\partial_\mu
  a\partial_\mu a^\dagger+\partial_\mu b\partial_\mu
  b^\dagger)]$, and the functional integral is well defined. Note
  that the symplectic fermion part of the Boltzmann weight is of the 
  form $\exp[-|cst|\int (\partial_\mu\eta
  \partial_\mu \eta^\diamond+\eta\eta^{\diamond}\partial_\mu \eta\partial_\mu
  \eta^\diamond)]$, and also exhibits the same sign as the action of
  the supersphere sigma model in the massive phase (where the symmetry is
  restored).

The exact S matrix can be deduced from the TBA by noticing that, for the matrix $S\otimes S$, the 
presence of the self coupling for the first node in the sigma model TBA would lead to a 
{\sl double} self coupling. This has to be removed, and the
usual calculation
gives 
\begin{equation}
    S_{PCM}=  Y~S_{\sigma}\otimes S_{\sigma}
\end{equation}
where the CDD factor $\hat{Y}={\sinh \omega+\sinh 2\omega\over \sinh
  3\omega}$,
$Y={\sinh\theta+i\sin(\pi/3)\over\sinh\theta-i\sin(\pi/3}$ cancels the
 double poles and double zeroes in $\Sigma_0^2$ (\ref{sigma0elt}). Let us recall for completeness the sigma model S matrix. 

\begin{equation}
\check{S}_{i_1j_1}^{j_2i_2}=\sigma_1 E+\sigma_2 P+\sigma_3 I \label{maini} 
\end{equation}
where  we have set

\begin{equation}
E_{i_1j_1}^{j_2i_2}=\delta_{i_1,\bar{j}_1}\delta^{i_2,\bar{j}_2}
(-1)^{x(i_1)}(-1)^{x(i_2)}
\end{equation}
while $P$ is the graded permutation operator 
\begin{equation}
P_{i_1j_1}^{j_2i_2}=(-1)^{p(i_1)p(j_1)}\delta_{i_1}^{i_2}\delta_{j_1}^{j_2}
\end{equation}
The indices $i$ take values in the fundamental representation
of the $osp(1/2)$ algebra, $i=1,2,3$. We set
$\bar{1}=1,\bar{2}=3,\bar{3}=2$, $x(1)=x(3)=0,x(2)=1$. The factors $\sigma$ in (\ref{maini}) read
\begin{eqnarray}
\sigma_1&=&-{2i\pi\over (N-2)(i\pi-\theta)}~\sigma_2\nonumber\\
\sigma_3&=&-{2i\pi\over (N-2)\theta}~\sigma_2\label{mainii}
\end{eqnarray}
for the value $N=1-2=-1$ characteristic of the $OSP(1/2)$ case.

\section{Realizations of the $UOSP(1/2)$ symmetry.}

In section 4, we have found two families of models  whose $S$ matrix 
has  $UOSP(1/2)$ symmetry
. The models 
based on the lattice TBA for $s$ integer correspond to 
$UOSP(1/2)_{k=s}$ WZW models peturbed by a current current 
interaction. The UV theory is a current algebra, in which the symmetry
is locally realized by two sets of currents, $J^{\pm,0},j^\pm$ and 
$\bar{J}^{\pm,0},\bar{j}^\pm$. 

What happens in the other family of models is less clear. 
An exception to this is the case $s=1/2$, ie the
$UOSP(1/2)/SU(2)\equiv OSP(1/2)/SP(2)$ 
sigma model. In this case, the symemtry is realized non linearly, 
and it is worthwhile seeing more explicitely how this works. 

\subsection{Symplectic fermions and non linearly realized symmetries}

Consider thus the supersphere sigma model. This model for positive
coupling 
describes the Goldstone phase for  
  $OSP(1/2)$ symmetry  broken down spontaneously to $SP(2)$
 (possible since the group is not unitary compact). 
For negative coupling, it is massive,
and the $OSP(1/2)$ symmetry is restored at large distance.
In either case, the action is proportional to (we have
slightly changed the normalizations compared with the previous paper)
\begin{equation}
S\propto 2\partial_\mu\eta_1\partial_\mu\eta_2+ (\partial_\mu x)^2
\end{equation}
with $2\eta_1\eta_{2}+x^{2}=1$. We can find the Noether currents with 
the usual procedure. An infinitesimal
$OSP(1/2)$ transformation reads
\begeqar
\delta x&=&-\delta\xi_{1}\eta_{1}+\delta\xi_{2}\eta_{2}\nonumber\\
\delta \eta_{1}&=&-\delta\xi_{2}x+\delta a\eta_{1}+\delta 
c\eta_{2}\nonumber\\
\delta \eta_{2}&=&-\delta\xi_{1}x+\delta b\eta_{1}-\delta a\eta_{2}
\endeqar
where $\delta\xi_{1},\delta\xi_{2}$ are `small' fermionic deformation 
parameters, $\delta a,\delta c$ small
bosonic parameters. By definition, this change leaves
$2\eta_{1}\eta_{2}+x^{2}$ invariant. In terms 
 of the fermion variables, the symmetry is realized non linearly:
\begeqar
\delta\eta_{1}=-\delta\xi_{2}(1-\eta_{1}\eta_{2})+\delta 
a\eta_{1}+\delta c\eta_{2}\nonumber\\
\delta\eta_{2}=-\delta\xi_{1}(1-\eta_{1}\eta_{2})+\delta 
a\eta_{1}-\delta a\eta_{2}
\endeqar

Performing the change in the action, and identifying 
the coefficients of linear derivatives 
$\partial_{\mu}\eta_{i},\partial_{\mu}x$ with the currents
gives five conserved currents. Three of them generate the sub
$su(2)$:
\begeqar
J^{+}&=&-{1\over 2}\eta_{1}\partial\eta_{1}\nonumber\\
J^{-}&=&{1\over 2}\eta_{2}\partial\eta_{2}\nonumber\\
J^3&=&{1\over 4}\left(\eta_{1}\partial\eta_{2}-\partial\eta_{1}\eta_{2}\right)
\endeqar
%
%
The two fermionic currents meanwhile are
\begeqar
j^+=\partial_{\mu}x~\eta_{1}-x~\partial_{\mu}\eta_{1}=
\partial_{\mu}\eta_{1}(1-2\eta_{1}\eta_{2})\nonumber\\
j^-=\partial_{\mu}x~\eta_{2}-x~\partial_{\mu}\eta_{2}=\partial_{\mu}\eta_{2}(1-2\eta_{1}\eta_{2})
\endeqar
These five currents should be present in the UV limit of 
the sigma model, which coincides with 
symplectic fermions. The latter  theory has been studied a great
deal. Of particular interest is the operator content, which is
conveniently encoded in the generating function (\ref{parfun}).
Recall that
the ``ground state'' (that is, fields of weight $(0,0)$) is degenerate
four times, while there are eight fields of weight $(1,0)$ (and eight
fields of weight $(0,1)$). It has sixteen fields of weight $(1,1)$. 
We can understand these multiplicities
easily by using the sigma model interpretation. 
From the $OSP(1/2)$ symmetry, 
we expect to have, by taking the weak coupling limit of the foregoing 
currents, five fields $(1,0)$ and five fields $(0,1)$ 
(these fields are not chiral currents, because of some logarithmic
festures:
more about this below). Meanwhile,
the broken $OSP(1/2)$ symmetry implies the existence of three non 
trivial fields with weight $(0,0)$, whose derivatives are also 
necessarily `currents'. We therefore 
expect {\sl eight} fields ($8=$ fundamental $+$ adjoint) $(1,0)$ and $(0,1)$,
in agreement with the known result. 

Note that fields with weights $(1,0)$ and $(0,1)$ can have some common
components
due to the presence of fields with vanishing weights. It follows that
many of their products do actually vanish, leading to a multiplicity
of sixteen for fields $(1,1)$, and not $8^2$, as one could have
naively assumed.

An interesting question is now what remains of the $OSP(1/2)$ symmetry
{\sl right at the weak coupling fixed point}, that is, in the
symplectic fermions theory itself. There, it turns out that only the
sub $SP(2)$ can be observed, as the bosonic currents $J^\pm, J^3$ are
still conserved in the symplectic fermion theory. This conservation 
boils down to the equations of motion 
$\partial_{\mu}\partial^{\mu}\eta_{i}=0$. If one naively 
tries to check the conservation of the fermionic currents, it seems 
one needs 
$\partial_{\mu}\partial^{\mu}(\eta_{1}\eta_{2})=0$, which is manifestly
wrong! So these currents, which are conserved in the sigma model at 
any
non zero coupling, are not strictly speaking conserved right at the
weak coupling fixed point.  

The explanation  of this apparent  paradox lies in the role of the 
coupling constant and how exactly one can obtain the conformal
limit. The best is to take the Boltzmann weight as $e^{-S}$ with $S$ 
as above,
$
S=2\partial_\mu\eta_1\partial_\mu\eta_2+ (\partial_\mu x)^2
$
and put the coupling constant in the radius of the supersphere 
$x^{2}+2\eta_{1}\eta_{2}=g^{2}$,
which now leads to  $x=g-{1\over g}\eta_{1}\eta_{2}$. The equations of
motion are
\begeqar
\partial_\mu\partial^\mu~x&=&\lambda x\nonumber\\
\partial_\mu\partial^\mu~\eta_1&=&\lambda \eta_1\nonumber\\
\partial_\mu\partial^\mu~\eta_2&=&\lambda \eta_2
\endeqar
where 
\begin{equation}
\lambda={1\over
g^2}\left[x\partial_\mu\partial^\mu~x+\eta_1\partial_\mu\partial^\mu\eta_2+\partial_\mu\partial^\mu\eta_1\eta_2\right]
\end{equation}
leading, as usual, to the conservation of $j^\pm$. 
The conformal symplectic fermion theory is then
obtained in the (singular) limit $g\rightarrow\infty$, where the field $x$
formally becomes a constant, and $\partial_{\mu}\partial^{\mu}x=0$ a
triviality. 
Within this limit, the $OSP(1/2)$ symmetry is lost, but one gets as
its remnant the  two fermionic ``currents'', $\partial_\mu\eta_1$ and
$\partial_\mu\eta_2$. 

It is interesting finally to
 discuss the algebra satisfied by the $SP(2)$ currents right at the
 conformal point (a related calculation has been presented in
 \cite{Kogan}, but we do not think its interpretation - based on
 rescaling the currents- is appropriate). 
 The OPE's are rather complicated:
\begeqar
J^{+}(z)J^{-}(w)&=&{1\over 4}{1+2\ln|z-w|+\eta_{2}\eta_{1}\over(z-w)^{2}}
-{1\over 8} {\partial(\eta_{1}\eta_{2})
+{\bar{z}-\bar{w}\over z-w}\bar{\partial}(\eta_{1}\eta_{2})\over z-w}
-{1\over 2}\ln|z-w|\partial\eta_{1}\partial\eta_{2}\nonumber\\
&+&{{3\over 2}J^3+{1\over 
2}{z-w\over\bar{z}-\bar{z}}\bar{J}^3\over z-w}\\
J^3(z)J^{\pm}(w)&=&\pm {{3\over 4}J^{\pm}+{1\over 
4}{\bar{z}-\bar{w}\over z-w}\bar{J}^{\pm}
\over z-w}\nonumber\\
J^3(z)J^3(w)&=&{1\over 8}{1+2\ln|z-w|+\eta_{2}\eta_{1}\over(z-w)^{2}}
-{1\over 16} {\partial(\eta_{1}\eta_{2})
+{\bar{z}-\bar{w}\over z-w}\bar{\partial}(\eta_{1}\eta_{2})\over z-w}
-{1\over 4}\ln|z-w|\partial\eta_{1}\partial\eta_{2}\nonumber
\endeqar
and we see that the notation $J(z)$ is abusive: the field 
has weights $(1,0)$ but the OPEs involve $\ln \bar{z}$ terms. 
The commutators of charges are only affected by the ${1\over z-w}$ 
term, and the $su(2)$ relations 
are recovered not through a rescaling but because of the presence of 
other non trivial OPEs between
the `left' and `right' components. For instance, writing only the 
relevant term, one has
\begeqar
J^{+}(z)\bar{J}^{-}(w)&=&{1\over 2(z-w)}\bar{J}^3+{1\over 
2(\bar{z}-\bar{w})}J^3\nonumber\\
\bar{J}^{+}(z)\bar{J}^{-}(w)&=&{1\over 2(z-w)}\bar{J}^3+{1\over 
2(\bar{z}-\bar{w})}J\nonumber\\
\bar{J}^{+}(z)\bar{J}^{-}(w)&=&{{3\over 2}\bar{J}^3\over \bar{z}-\bar{w}}
\endeqar
ensuring $[Q^{+},Q^{-}]=2Q^3$, where $Q={1\over 2i\pi}\int (Jdz-\bar{J}d\bar{z})$.  

Amusingly, the $1/(z-w)^2$ part of the OPEs corresponds to the
normalization $k={1\over 2}$, so the UV limit of the sigma model does
contain a ``logarithmic $k=1/2$'' $su(2)$ current algebra.

\subsection{Speculations on the $SU(2)_{k}\otimes UOSP(1/2)/SU(2)$.}

It is tempting to speculate then 
that the models for $s$ half integer correspond to ``higher level'' generalizations of the symplectic fermions, with
a non linear realization of the $UOSP(1/2)$ symmetry, and  
a ``logarithmic  $su(2)$ current algebra''. We do not know what the action of these models might
 be, except that in the UV they should reduce to the tensor product
 of a $SU(2)_k$ WZW model and symplectic fermions. Notice of course  that 
 the PCM model - the limit $k\rightarrow\infty$, does obey this scenario.   
  Indeed, the PCM model also provides a  realization of the $UOSP(1/2)$
symmetry which is non linear once the constraints 
have been explicitely solved. Solving the constraints in terms of the
fermions gives 
\begin{eqnarray}
    J^+&=&{1\over 2}
    \left[b\partial a^{\diamond}-a^{\diamond}\partial b-{b^2\over
      4}\eta^\diamond\partial\eta^\diamond {(a^{\diamond})^2\over
      4}\eta\partial\eta-{ba^{\diamond}\over
     4}\left(\eta\partial\eta^\diamond+\eta^\diamond\partial\eta\right)
     \right]
      \nonumber\\
J^-&=&{1\over 2}\left[a\partial b^{\diamond}-b^{\diamond}\partial a+{a^2\over
  4}\eta^\diamond\partial\eta^\diamond +{(b^{\diamond})^2\over
  4}\eta\partial\eta-{ab^{\diamond}\over
  4}\left(\eta\partial\eta^\diamond+\eta^\diamond\partial\eta\right)
  \right]
\nonumber\\
J^3&=&2\left[a\partial a^{\diamond}+b^{\diamond}\partial b
+{1\over
4}\left(ab\eta^\diamond\partial\eta^\diamond-a^{\diamond}
b^{\diamond}\eta\partial\eta\right)+{aa^{\diamond}\over 
8}\left(\partial\eta\eta^\diamond-\eta\partial\eta^\diamond\right)\right.\nonumber\\
\left.~~~+{bb^{\diamond}\over 
8}\left(\eta\partial\eta^\diamond-\partial\eta\eta^\diamond\right)\right]
\end{eqnarray}
The fermionic currents meanwhile read \footnote{It is useful to
recall that   factoring out the $SU(2)$, ie taking as action $j^{+}j^{-}$, 
leads  (after some rescalings and relabellings) to the 
action of the supersphere sigma model $UOSP(1/2)/SU(2)$ written
earlier in terms of  $\eta_{1},\eta_{2}$. }
\begin{eqnarray}
    j^+&=&-{1\over 2}\left(b\partial\eta^\diamond+a^{\diamond}\partial\eta\right)
    -{b\over 16}\eta\eta^\diamond\partial\eta^\diamond-{a^{\diamond}\over
      16}\eta\eta^\diamond\partial\eta\nonumber\\
    j^-&=&-{1\over 2}\left(b^{\diamond}\partial\eta-a\partial\eta^\diamond\right)
+{1\over 16}a\eta\eta^\diamond\partial\eta^\diamond-{b^{\diamond}\over
  16}\eta\eta^\diamond\partial\eta
\end{eqnarray}
One can as well solve for the bosonic constraint $aa^{\diamond}+
bb^{\diamond}=1$. If
one does so, and rescales the fields with the coupling constant as in
the supersphere case, the UV expression of the currents becomes simply
the sum of the currents for a system of 3 bosons (the small coupling
limit of the $SU(2)$ PCM model) and the currents for the symplectic
fermion theory.

The evidence from the TBA is that the PCM model can give rise to two 
kinds of models (more on this in the conclusion): either the $UOSP(1/2)_{k}$ WZW models like in the
usual case, but also the $SU(2)_{k}\otimes UOSP(1/2)/SU(2)$ model,
which presumably involves some sort of term changing the $SU(2)$ part
of the action into the WZW one with a current current perturbation, 
but leaving the symplectic fermionic part essentially unaffected. We 
do not know how to concretely realize this though.

Another interesting aspect stems from the fact that the central charge obtained by giving antiperiodic  boundary conditions
to the kinks reads,
after elementary algebra, 
\begin{equation}
c=1-6{(k+1)^2\over (k+2)}
\end{equation}
This is precisely the central charge of the models $M_{k+2,1}$, of which the
first two have  $c=-1$ and $c=-7$. We are thus led to speculate that 
the $M_{k+2,1}$ models - or rather, their proper `non minimal'
versions (studied in \cite{Flohr}, although we do not necessarily
agree with the conclusions there), as the minimal models are entirely empty in this
case, are models with spontaneously broken $UOSP(1/2)$ symmetry. It
would be very interesting to look further for signs of an $UOSP$
structure in these models, and to study their `logarithmic' $SU(2)$
algebra. 

Note that these models are 
obtained 
by hamiltonian reduction of the $SU(2)_k$ model. In this reduction 
\cite{bershoog}, an
auxiliary $\eta\xi$ system is introduced to play the role of
Fadeev-Popov ghosts, so these models are indeed naturally related to
the product of $SU(2)_k$ and $U(1)$ as we observed earlier.

\section{The $UOSP(1/2)/U(1)$ sigma model(s)}

Instead of factoring out the $SU(2)$, one can of course
also factor out the $U(1)$ and get an $UOSP(1/2)/U(1)$ sigma model. 
This is especially interesting since the standard argument to derive 
the continuum limit of the $osp(1/2)$ spin chains would lead to a sigma 
model on the manifold parametrizing the coherent states, and this is
precisely $UOSP(1/2)/U(1)$ \cite{Balantekin,Gradechi}. 

Note however that the manifold $UOSP(1/2)/U(1)$  is not a symmetric (super)
space (this can easily be seen since the (anti) commutator of two
fermionic generators does not always belong to the Lie algebra of
$U(1)$). As a consequence, sigma models on this manifold 
will have more than one coupling constant.

To proceed, a possible strategy is to follow 
 \cite{FateevZamo} and consider for a while 
models $UOSP(1/2)_k/U(1)$, that is graded parafermionic theories. 

Graded parafermions \cite{Camino}  
theories are constructed in a way similar to the original construction of
 Fateev and Zamolodchikov, with the additional ingredient of a $Z_2$
 grading. They obey the OPE rules
\begin{equation}
\psi_l(z)\psi_{l'}(w)=(-1)^{p(l)p(l')} \exp\left(2i\pi {ll'\over
    k}\right) \psi_{l'}(w)\psi_l(z)
\end{equation}
Their dimensions are $h_l={l(k-l)\over k}+{\epsilon(l)\over 2}$, where
$\epsilon=1$, $l$ half an odd integer, $\epsilon=0$ otherwise. 
Of particular interest is
the OPE 
\begin{eqnarray}
\psi_{1/2}(z)\psi_{-1/2}(w)&=&(z-w)^{{1\over 2k}-2} \left[1+(z-w)^2
  O^{(1/2)}+\ldots\right]\nonumber\\
\psi_{1}(z)\psi_{-1}(w)&=&(z-w)^{{2\over k}-2} \left[1+(z-w)^2
  O^{(1)}+\ldots\right]\nonumber\\
\end{eqnarray}
Here, the operators $O$ have dimension 2, and must obey
$O^{(1)}-O^{(2)}={2k+3\over 2k} T$, $T$ the stress energy tensor. The 
simplest parafermionic theory for $k=1$ has $c=-{3\over 5}$, and
seems to coincide 
with the model $M_{5,3}$ \footnote{Since $SU(2)_1$ can be
represented in terms of a free boson, the cosets  $OSP(1/2)/SU(2)$ 
and $
OSP(1/2)/U(1)$ are equivalent there.}.  
For $k$ an integer, $l$ runs
over the set $l=-k+{1\over 2},\ldots,0,\ldots,k-{1\over 2}$, $2l\in
Z$. Parafermions with integer $l$ are bosonic, the others are
fermionic. For $k=1$, $\psi_1\equiv I$, and there is only a pair of
parafermionic fields, of weight $h={3\over 4}$. It can be shown that the
parafermionic theories just defined coincide with
$UOSP(1/2)_{k}/U(1)$ coset theories.

Like in the $SU(2)_{k}$ case, the $UOSP(1/2)_k$ model with 
a current-current pertubation can be written in terms of the graded
parafermions and a free boson $\phi$. It is then easy to find an
integrable anisotropic deformation
\begin{equation}
\psi_1\bar{\psi}_{1} e^{i\beta\sqrt{2\over k}\phi}+
\psi_{-1/2}\bar{\psi}_{-1/2}e^{-{i\over 
\sqrt{2k}}\beta\phi}\label{parafpert}
\end{equation}
(In the case $k=1$, the perturbation reads
$e^{-i\sqrt{2}\beta\phi}+\psi_{1/2}\bar{\psi}_{1/2}e^{{i\over
    \sqrt{2}}\beta\phi}$. )
The non local conserved currents  \cite{BL} are $\psi_{-1}e^{-i\sqrt{2\over
k}{\varphi\over \beta}}$ and $\psi_{1/2}e^{{i\over
\sqrt{2k}}{\varphi\over \beta}}$ (where $\varphi$ denotes the right component of 
$\phi=\varphi+\bar{\varphi}$). 
The TBA and S matrices are rather 
obvious: we take the same left part of the
diagram as for the $OSP(1/2)_k$ case, but replace the infinite right
tail by the ubiquitous, finite and anisotropic part discussed 
in our first paper. In the isotropic limit \cite{BL}
$\beta^{2}\rightarrow 1$, the RG generates the other terms necessary 
to make (\ref{parafpert}) into a whole current current perturbation.

\begin{figure}
\begin{center}
\epsfig{file=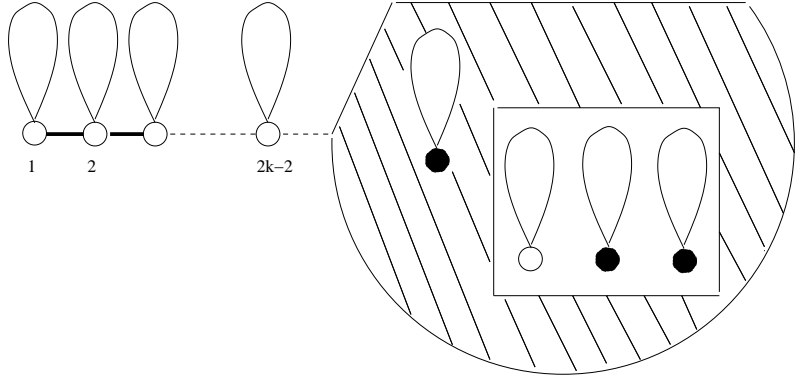,scale=0.5}  
\caption{\label{fig8} Conjectured incidence diagram of the TBA describing the $UOSP(1/2)_k/U(1)$  
parafermions.}
\end{center}
\end{figure}

Taking the limit $\beta\rightarrow 0$ would then lead to the TBA for
the parafermionic theory. This would require an understanding of the 
scattering in the attractive regime where bound states exist, but we 
have not performed the related analysis. It is possible however 
to make a simple conjecture based on numerology, and analogies with 
the $SU(2)$ case. Consider indeed the  TBA in Figure \ref{fig8}

\noindent where the box represents the set of couplings discussed in our first paper \cite{SWKI}. 
In the UV, the diagram is identical to the one arising in the study of
the $a_2^{(2)}$ Toda theory. The central charge is $c_1=2k-1$ as discussed in
\cite{SWKI}. In the IR, the diagram is identical to the ones
arising in the $UOSP/SU$ coset models, and $c_2=2k-4+{12\over
  2k+3}$. The final central charge is thus $c_{eff}=3-{12\over 2k+3}$,
and concides with the effective central charge for $UOSP(1/2)/U(1)$
parafermions of level $k$. We conjecture this TBA describes the
perturbation of these parafermionic theories by the combination of
graded parafermions
\begin{equation}
\psi_1\bar{\psi}_1+\psi_{-1/2}\bar{\psi}_{-1/2}
\end{equation}
The
 effective dimension of the perturbation deduced from the TBA is
 $1-{1\over 2k}$, and this coincides with the combination
 $h={2h_1+h_2\over 3}$. Note that we have not studied what kind of scattering theory would give rise to the TBA in Figure \ref{fig8}, and whether it is actually meaningful. Still,
taking the limit $k\rightarrow\infty$, we should obtain
the TBA for something that looks like an $UOSP(1/2)/U(1)$ sigma model.
Notice that the bosonic part of this model is identical with the
$SU(2)/U(1)$ sigma model,  and thus there is the possibility of a
topological term. It is not clear what the 
low energy limit of the model with topological angle $\theta=\pi$
would be.

\section{Conclusions}

The results presented here presumably  have rather simple generalization to the $OSP(1/2n)$ case, even though 
details might not be absolutely straightforward to work out - for instance, we do not know
of embeddings generalizing the one discussed in the first sections. 

The supersphere sigma model for $g$ positive in the conventions of
section 2, flows in the IR to weak coupling, at least 
perturbatively. It is expected that the phase diagram will
exhibit a critical point at some value $g^{*}$ and that for larger coupling, 
the theory will be  massive. The
critical point presumably coincides with the dilute $O(N=-1)$ theory 
first solved by Nienhuis \cite{Nienhuis}. This theory is 
described by a free boson with a charge at infinity, and is 
 closely related with the minimal model  
$M_{5,3}$. In fact  the partition function 
of the  dilute $O(N=-1)$ model provided one restricts to {\sl
  even} numbers of non contractible loops can be
written in the Coulomb gas language of Di Francesco et al. \cite{US} as
\begin{equation}
Z_{5,3}={1\over 2}\left[Z_c(3/5,5)-Z_c(3/5,1)\right]
\end{equation}
and coincides with the partition function of the minimal model. Earlier in this paper, we have 
identified this model 
with the $UOSP(1/2)_{1}/U(1)$ parafermionic theory.  The full $O(N=-1)$ theory,
however defined, has a considerably more complex 
operator content \cite{ReadS}.

Note that antiperiodic boundary conditions for the fermions, which give an effective 
central charge equal to 
$c_{eff}=1$ in the supersphere sigma model give, in the critical theory, a highly 
irrational value $c_{eff}=1-{18\over\pi^2}\left(\hbox{arccosh}(3/2)\right)^2$. There are no indications 
that an integrable flow from the critical theory to the low temperature generic theory exists. 
An integrable flow is known to exist in the special case where the symmetry is
enhanced to $SU(1/2)$. In that case,
the IR theory is the so called dense $O(N=-1)$ model, which has $c=-7$, and is closely related
with the minimal model $M_{3,1}$. Note that this model is the second model of 
the unidentified series in section 4, and bears some formal 
resemblance to the model $UOSP(1/2)_{3/2}$. What this means remains
one of the many open questions in this still baffling area.

\bigskip
\noindent {\bf Acknowledgments:} We thank G. Landi, N. Read   and M. Zirnbauer
for useful remarks and suggestions. We especially  thank P. Dorey for pointing out the discussion of $OSP$ coset models in \cite{Dorey}. The work of HS was supported in
part by the DOE.

\appendix

\section{Some results on $osp(1/2)$.}

We collect in this appendix some formulas about $osp(1/2)$, the
associated current algebra and groups.

The supergroup  $OSP(1/2)$ is the group of `real' matrices $g$ 
obeying (basic references are \cite{BereTol,scheunert,frappat})
\begin{equation}
g^{st}~J~g=J
\end{equation}
where  \footnote{For $g$ a bosonic matrix, 
$g=\left(\begin{array}{cc} 
a&b\\c&d\end{array}\right)$, recall
that $g^{st}=\left(\begin{array}{cc} 
a^t&c^t\\-b^t&d^t\end{array}\right)$.}
\begin{equation}
J=\left(\begin{array}{ccc}
1&0&0\\
0&0&1\\
0&-1&0\end{array}\right)
\end{equation}
Elements of the group preserve the quadratic form, if 
$X=(b,\theta_1,\theta_2)$,
$X.X'=bb'+\theta_1\theta_2'-\theta_2\theta_1'$. 
They can be parametrized by $g=e^A$ with
\begin{equation}
A=\left(\begin{array}{ccc}
0&-\eta_1&-\eta_2\\
\eta_2&a&c\\
-\eta_1&b&-a
\end{array}\right)
\end{equation}
Here no complex conjugation is ever needed: $a,b,c$ are real numbers, 
and $\eta_1,\eta_2$ are `real'
Grassman numbers. 

The group $UOSP(1/2)$ in contrast is  made of complex
supertransformations satisfying 
\begin{equation}
g^{st}~J~g=J,~~g~g^\ddag=1
\end{equation}
To define the adjoint $M^\ddag$, we first need to introduce 
a complex conjugation denoted by $\diamond$. It is, technically, a 
graded involution, which coincides with complex
conjugation for pure complex numbers, $c^\diamond=\bar{c}$, $c\in C$,
and obeys in general \footnote{Recall that it is not possible to 
define a unitary version of $OSP$
with
the usual conjugation.}
\begin{eqnarray}
(xy)^\diamond=x^\diamond y^\diamond\nonumber\\
(x^\diamond)^\diamond=(-)^{p(x)} x\nonumber\\
(cx)^\diamond=\bar{c} x^\diamond
\end{eqnarray}
One then sets $g^\ddag=\left(g^{st}\right)^\diamond$
\footnote{Recall that the $\ddag$ operation obeys the usual properties,
$(hh')^\ddag = (h')^\ddag h^\ddag$. It can be considered as the combination
of the $\dagger$ operation in the Lie algebra (see the appendix), and the $\diamond$ operation on `scalars'.} 
, so $g$ in
$UOSP(1/2)$ preserves in addition the form 
$X^\diamond
X'=\bar{b}b'+\theta_1^\diamond\theta_1'-\theta_2^\diamond\theta_2'$. 

One has now 
 $g=e^A$ with
\begin{equation}
A=\left(\begin{array}{ccc}
0&-\eta&-\eta^\diamond\\
\eta^{\diamond}&ia&ib\\
-\eta&ib^\diamond&-ia
\end{array}\right)
\end{equation}
with $a$ real, $a^\diamond=a$. The fermionic content of the supergroup is
essentially unchanged, with $\eta\equiv \eta_1$, $\eta^\diamond\equiv
\eta_{2}$. But the bosonic content is different: the non compact
bosonic subgroup $SP(2)$ has been replaced by the compact one
$SU(2)$. 

The  algebra $osp(1/2)$ is generated  by operators 
which we denote $J^{3},J^\pm$ (bosonic) and $j^\pm$ (fermionic). Their commutation relations can be obtained from the current algebra given below by restricting to the zero modes. The 
casimir reads
\begin{equation}
C=(J^{3})^2+{1\over 2}\left(J^+J^-+J^-J^+\right)+{1\over
  4}\left(j^-j^+-j^+j^-\right)
\end{equation}
The representations of the super Lie algebra are labelled by an
integer or half integer $j$, and are of dimension $4j+1$.  The
fundamental representation is three dimensional, and has spin $j=1/2$. It does
contain a sub $sl(2)$ fundamental representation, following the
pattern of $J^{3}=\hbox{diag}(0,-1/2,1/2)$. The generators
$J^\pm,J^{3}$
are bosonic. The fermionic generators 
are given by 
\begin{equation}
j^+=\left(\begin{array}{ccc}
0&-1&0\\
0&0&0\\
-1&0&0\end{array}\right)
~~~
j^-=\left(\begin{array}{ccc}
0&0&-1\\
1&0&0\\
0&0&0\end{array}\right)
\end{equation}

The only metric compatible with $osp(1/2)$ requires the definition of
a generalized adjoint satisfying  (here $p=0,1$ denotes the parity)
\cite{scheunert}
\begin{equation}
\left<A^\dagger \alpha|\beta\right>=(-1)^{p(A)p(\alpha)}
\left<\alpha|A\beta\right>
\end{equation}
and thus 
\begin{equation}
(AB)^\dagger =(-1)^{p(A)p(B)} B^\dagger A^\dagger
\end{equation}
It follows that $(J^\pm)^\dagger=J^\mp$, $(J^{3})^\dagger=J^{3}$, while
there remains some freedom for the fermionic generators,
$(j^+)^\dagger =\pm j^-$, $((j^+)^\dagger )^\dagger =-j^+$. It is in 
the nature of the algebra that 
negative norm square  states will  appear whatever the choice. Indeed, let 
us choose for instance 
\begin{equation}
\left(j^{+}\right)^\dagger=j^{-},\left(j^{-}\right)^\dagger=-j^{+}
\end{equation}
It then follows that the norm square of the state $|j,m>$ is 
\begin{equation}
<j,m|j,m>=(-)^{2p(j)(j-m)}
\end{equation}
Here, $p(j)=0$ if the highest weight state $|j,j>$ is bosonic,
$p(j)=1$ if it is fermionic. Even if we start with the fundamental
representation $j=1/2$ with $|1/2,1/2>$ bosonic, in the tensor 
product of
this representation with itself, representations where the highest
weight is fermionic will necessary appear. These do contain negative
norm square states. In this paper, we will always choose the
gradation for which $|1/2,1/2>$ is fermionic, and thus the fundamental
representation
has superdimension equal to $-1$. 

The 
current algebra is defined by 
\begin{eqnarray}
\left[J^3_n,J^\pm_m\right]=\pm J^\pm_{n+m} &
\left[J_n^3,J_m^0\right]={k\over 2} n\delta_{n+m}\nonumber\\
\left[J_n^+,J_m^-\right]&=kn\delta_{n+m}+2J^3_{n+m}\nonumber\\
\left[J_n^3,j_m^\pm\right]=\pm {1\over 2}j^\pm_{m+n} & 
\left[J_n^\pm,j_m^\pm\right]=0\nonumber\\
\left[J_n^\pm,j_m^\mp\right]=-j_{n+m}^\pm &
\left\{j_n^\pm,j_m^\pm\right\}=\pm 2J_{n+m}^\pm\nonumber\\
\left\{j_n^+,j_m^-\right\}&=2kn\delta_{n+m}+2J_{n+m}^3
\end{eqnarray}
Normalizations are such that the algebra contains a sub $sl(2)$
current algebra at level $k$.

The Wess Zumino Witten model on the supergroup $UOSP(1/2)$ corresponds
to $k$ positive  integer, and the sub $sl(2)$ current algebra to the 
WZW
model  $SU(2)_k$.

As commented in the text, the supersphere $S^{3,2}$ is the
supermanifold of the supergroup $UOSP(1/2)$. It is also the total
space of a principal fibration with structure group $U(1)$ and the
quotient of this action is just the supersphere $S^{2,2}\approx
UOSP(1/2)/U(1)$. The explicit realization is as follows \cite{Landi}. 
Setting
\begin{eqnarray}
x_0&=&(aa^\diamond-bb^\diamond)\left(1-{1\over
  4}\eta\eta^\diamond\right)\nonumber\\
x_1&=&(ab^\diamond+ba^\diamond))\left(1-{1\over
  4}\eta\eta^\diamond\right)\nonumber\\
x_2&=&i(ab^\diamond-ba^\diamond)
)\left(1-{1\over
  4}\eta\eta^\diamond\right)\nonumber\\
\eta_1&=&-{1\over 2}(a\eta^\diamond+\eta b^\diamond)\nonumber\\
\eta_2&=&{1\over 2}(\eta a^\diamond-b\eta^\diamond)\label{sphparam}
\end{eqnarray}
(these obey $x_i^\diamond=x_i$, and
$\eta_1^\diamond=-\eta_2$) we obtain points in $S^{2,2}$, since 
$\sum(x_{i})^{2}+2\eta_{1}\eta_{2}=1$. 
Conversely, for a given point $x_{0},x_{1},x_{2},\eta_{1},\eta_{2}$
of $S^{2,2}$
one gets
\begin{eqnarray}
{1\over 2}\eta\eta^\diamond&=&\eta_1\eta_2\nonumber\\
aa^\diamond&=&{1\over 2}\left[1+x_0(1+{1\over 
2}\eta_1\eta_2)\right]\nonumber\\
bb^\diamond&=&{1\over 2}\left[1-x_0(1+{1\over 
2}\eta_1\eta_2)\right]\nonumber\\
ab^\diamond&=&{1\over 2}(x_1-ix_2)(1+{1\over 
2}\eta_1\eta_2)\nonumber\\
\eta a^\diamond&=&-(x_1+ix_2)\eta_1+(1+x_0)\eta_2\nonumber\\
\eta b^\diamond&=&(x_1-ix_2)\eta_2-(1-x_0)\eta_1
\end{eqnarray}
Define finally $U(1)=\{ w,w\hbox{ bosonic }, ww^\diamond=1\}$. Since 
the
 parametrization of (\ref{sphparam}) is invariant under
 $(a,b,\eta)\rightarrow (wa,wb,w\eta)$, this proves the statement. 
 
 Of course, the two spaces
 $UOSP(1/2)/U(1)$ and $S^{2,2}$
 are not  topologically equivalent: the fibration just discussed 
 is in fact a `superextension' of the Dirac monopole \cite{Landi}.

\end{document}